\documentstyle[12pt,twoside,fleqn,espcrc1,psfig]{article}
\newcommand{\AmS}{{\protect\the\textfont2
  A\kern-.1667em\lower.5ex\hbox{M}\kern-.125emS}}

\title{Kaon and Hyperon Form Factors in Kaon Electroproduction on the Nucleon}

\author{T. Mart\address{Jurusan Fisika, FMIPA, 
        Universitas Indonesia, Depok 16424, Indonesia}$^{\rm ,b}$,
        C. Bennhold\address{Center for Nuclear Studies,
        Department of Physics, The George Washington
        University, $~\!~$ Washington, D.C. 20052, USA}}

\begin{document}
\maketitle

\begin{abstract}
The electromagnetic form factors of strange mesons and baryons are
studied by means of kaon electroproduction on the nucleon. The response 
functions that are sensitive to the $K^0$, $\Lambda$, $\Sigma^+$, and 
$KK^*\gamma$ transition form factors are systematically explored.
The effects of these form factors on several response functions
are discussed.
\end{abstract}

\section{INTRODUCTION}
The study of the electromagnetic structure of hadrons provides an important
key for our understanding of hadronic structure, since form factors
reflect the charge distribution of quarks and gluons inside the
hadron. During the last decades there has
been considerable effort to develop models for not only the nucleon, but also 
hyperon and meson form factors. Nevertheless, 
the main problematic issue is the lack of experimental verification of these
models due to the lack of stable targets in the case of strange hadrons. 

Unlike the case of the proton, where both electric and magnetic form
factors [$G_E(q^2)$ and $G_M(q^2)$] can be extracted directly, the
measurement of charged meson (e.g. $\pi^+$ and $K^+$) form factors
requires an indirect technique. The method is known as Chew-Low extrapolation 
 \cite{chew-low,baker93} and based on the extrapolation of
the electroproduction data to the pion/kaon pole, where the dominance of
the longitudinal differential cross section can isolate the contribution
of the pion/kaon form factor from the contamination of other diagrams.
However, there is no certainty that such techniques are also applicable 
to other strange mesons and hyperons (e.g., $K^0$, $\Lambda$ and $\Sigma$),
since the required experimental data are not available at the present. 

In this paper, we suggest that kaon electroproduction may provide at least
an indirect method for obtaining information on these form factors. The 
result presented here provides an extension of our previous report
\cite{trieste97}.

\section{RESPONSE FUNCTIONS IN KAON ELECTROPRODUCTION}
While the cross section for an experiment using unpolarized electron beams has
only four individual terms (see e.g. Ref.~\cite{david}), the differential
cross section for kaon production using polarized electron, target, and
recoil may be written in terms of response functions as \cite{knoechlein}

\begin{table}[!ht]
\caption{Complete response functions for pseudoscalar meson
         electroproduction~\protect\cite{knoechlein}. The polarization of
         the target (recoil) is indicated by $\alpha$ ($\beta$). The last
         three columns ($^cTL'$, $^sTL'$, and $TT'$) are response functions 
         for polarized electrons. $\ddagger$ denotes a response function 
         which does not vanish but is identical to another response function.}
\renewcommand{\arraystretch}{1.4}
\begin{center}
\begin{tabular}{|l|cc|cccccc|ccc|}
\hline\hline
Note & $\beta$ & $\alpha$ & $T$ & $L$ & $^cTL$ & $^sTL$ & $^cTT$ & 
 $^sTT$ & $^cTL'$ & $^sTL'$ & $TT'$\\
\hline\hline
Unpolarized&0&0&$R_T^{00}$ & $R_L^{00}$ & $R_{TL}^{00}$ & 0 & $R_{TT}^{00}$ & 
 0 &  0 & $R_{TL'}^{00}$ & 0 \\
\hline
Polarized & 0&$x$ & 0 & 0 & 0 & $R_{TL}^{0x}$ & 0 &  $R_{TT}^{0x}$ & 
 $R_{TL'}^{0x}$ & 0 &  $R_{TT'}^{0x}$ \\
target & 0&$y$ & $R_T^{0y}$ & $R_L^{0y}$ & $R_{TL}^{0y}$ & 0 & $\ddagger$ & 
 0 & 0 & $R_{TL'}^{0y}$ & 0 \\
& 0&$z$& 0 & 0 & 0 & $R_{TL}^{0z}$ & 0 & $R_{TT}^{0z}$ & $R_{TL'}^{0z}$ & 0 & 
  $R_{TT'}^{0z}$ \\
\hline
Polarized & ~$x'$~&0& 0 & 0 & 0 & $R_{TL}^{x'0}$ & 0 &  $R_{TT}^{x'0}$ & 
 $R_{TL'}^{x'0}$ & 0 &  $R_{TT'}^{x'0}$ \\
recoil & $y'$&0& $R_T^{y'0}$ & $\ddagger$ & $\ddagger$ & 0 & $\ddagger$ & 
 0 & 0 & $\ddagger$ & 0 \\
&$z'$&0& 0 & 0 & 0 & $R_{TL}^{z'0}$ & 0 & $R_{TT}^{z'0}$ & $R_{TL'}^{z'0}$ &0& 
  $R_{TT'}^{z'0}$ \\
\hline
Polarized & $x'$&$x$& $R_T^{x'x}$ & $R_L^{x'x}$ & $R_{TL}^{x'x}$ & 0 & 
 $\ddagger$ & 0 &  0 & $R_{TL'}^{x'x}$ & 0 \\
target & $x'$&$y$&0&0&0&$\ddagger$&0&$\ddagger$&$\ddagger$&0&$\ddagger$\\
and & $x'$&$z$& $R_T^{x'z}$ & $R_L^{x'z}$ &$\ddagger$&0&$\ddagger$&0&0&$
 \ddagger$&0\\
recoil & $y'$&$x$&0&0&0&$\ddagger$&0&$\ddagger$&$\ddagger$&0&$\ddagger$\\
&$y'$&$y$&$\ddagger$&$\ddagger$&$\ddagger$&0&$\ddagger$&0&0&$\ddagger$&0\\
&$y'$&$z$&0&0&0&$\ddagger$&0&$\ddagger$&$\ddagger$&0&$\ddagger$\\
&$z'$&$x$& $R_T^{z'x}$ & $\ddagger$ & $R_{TL}^{z'x}$ & 0 & $\ddagger$ & 
 0 &  0 & $R_{TL'}^{z'x}$ & 0 \\
&$z'$&$y$&0&0&0&$\ddagger$&0&$\ddagger$&$\ddagger$&0&$\ddagger$\\
&$z'$&$z$&$R_T^{z'z}$&$\ddagger$&$\ddagger$&0&$\ddagger$&0&0&$\ddagger$&0\\
\hline\hline
\end{tabular}
\label{tab:response}
\end{center}
\end{table}

\begin{eqnarray}
\frac{d\sigma_v}{d\Omega_K}\!\!\!\! &&= ~\frac{\mid {\vec q}_K 
\mid}{k_{\gamma}^{\rm cm}} P_{\alpha} P_{\beta} \left\{R_T^{\beta \alpha}
+ \varepsilon_L R_L^{\beta \alpha } + \left[ 2 \varepsilon_L \left( 1 +
\varepsilon \right) \right]^{\frac{1}{2}} ( ^c \! R_{TL}^{\beta \alpha}\cos 
\phi_{K}+~\! ^s\! R_{TL}^{\beta \alpha} \sin \phi_{K} )\right.\nonumber\\
&& 
+~ \varepsilon ( ^c \! R_{TT}^{\beta \alpha} \cos 2 \phi_{K}
+~\! ^s\! R_{TT}^{\beta \alpha} \sin 2 \phi_{K} ) + 
 h \left[ 2 \varepsilon_L ( 1 - \varepsilon ) \right]^{\frac{1}{2}}
 ( ^c \! R_{TL'}^{\beta \alpha} \cos \phi_{K}
+~\! ^s\! R_{TL'}^{\beta \alpha} \sin \phi_{K} ) \nonumber \\
&& 
\left. +~ h (1-\varepsilon^2 )^{\frac{1}{2}} R_{TT'}^{\beta \alpha} \right\}, 
\end{eqnarray}
where $P_{\alpha} = (1, P_x, P_y, P_z)$ and $P_{\beta} = (1, P_{x'},
P_{y'}, P_{z'})$ indicate the target and the recoil polarization vectors, 
respectively. In total, there are 36 different response functions in the 
electroproduction of pseudoscalar mesons as listed in Table \ref{tab:response},
although it is not necessary to measure them all for a complete experiment.

\section{THE ELEMENTARY OPERATOR}
The presently existing elementary models were mostly developed from fits to 
experimental data by taking a number of resonances and leaving their coupling 
constants as free parameters. However, the limited data base permits only
qualitative conclusions until now. For our purpose, we will employ the
elementary models of Ref.~\cite{williams} and use the methods of
Ref.~\cite{terry} to extend these models to the $n(e,e'K^0)\Lambda$
and $p(e,e'K^0)\Sigma^+$ channels. Since at present no data are available in 
both processes, it is impossible to test the reliability of the models in
these isospin channels. Therefore, for the present we are only able to
estimate the relative effects that form factors have on particular
response functions.

\section{THE ELECTROMAGNETIC FORM FACTORS}

\subsection{The $K^0$ Form Factor}
The existence of the $K^0$ form factor provides one of the unique properties
of the neutral kaon compared with other neutral SU(3) pseudoscalar mesons. The
difference between the strange and non-strange quark masses creates a
non-uniform charge distribution in the $K^0$. As a consequence, although its
total charge is zero, the $K^0$ has an electric form factor. Since the mass
difference is still smaller than the mass scale associated with confinement 
in Quantum Chromodynamics (QCD), $(m_s-m_d)<\Lambda _{\rm QCD}$, it could 
lead to a sensitive test of phenomenological models that attempt to
describe nonperturbative QCD.

\begin{figure}[!ht]
\begin{minipage}[htb]{75mm}
{\psfig{figure=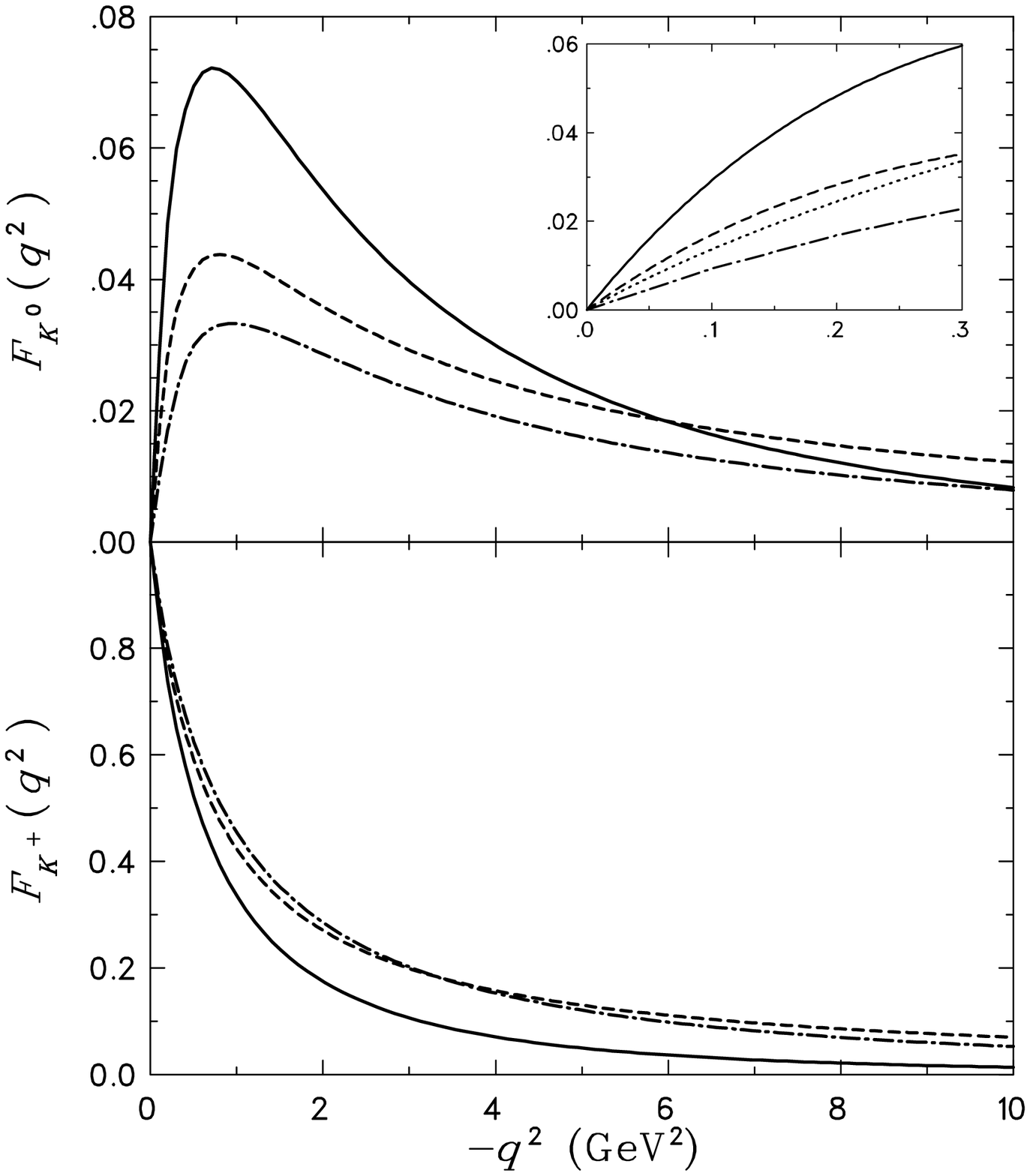,width=75mm}}
\caption{The $K^0$ and $K^+$ form factors predicted by the QMV calculation
(dash-dotted lines), VMD model (dashed lines), ChPT (dotted lines), and 
LCQ model (solid lines). The same units are used for the insert.}
\label{fig:ffk0}
\end{minipage}
\hspace{\fill}
\begin{minipage}[htb]{75mm}
{\psfig{figure=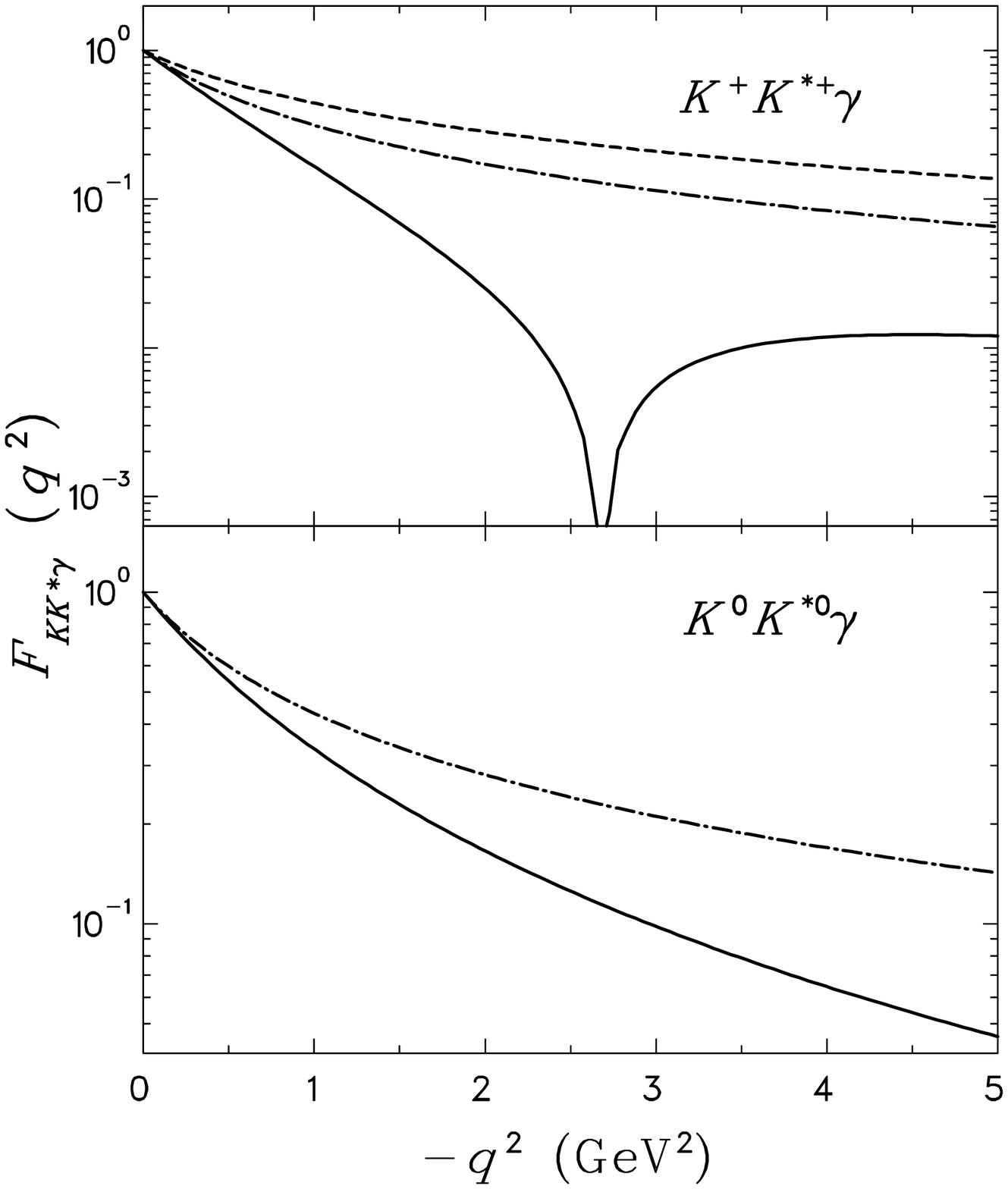,width=75mm}}
\caption{The transition form factors $K^+K^{*+}\gamma$ and $K^0K^{*0}\gamma$
 as predicted by Ref.~\protect\cite{williams} (dash-dotted lines) and
 Ref.~\protect\cite{muenz96} (solid lines). The dashed line shows the
 elastic $K^+$ form factor.}
\label{fig:fftr}
\end{minipage}
\end{figure}

In this paper we employ two relativistic quark models to calculate the
$K^0$ form factor, the light-cone quark (LCQ) model~\cite{ito1} and the
quark-meson vertex (QMV) model~\cite{buck}. The charge form factor of  
$K^0$  can be expressed as
\begin{eqnarray}
  F_{K^0}(q) &=& e_d F_L(q) + e_{\bar s}F_H(q) ~,
\label{eq:fk0}
\end{eqnarray}
where $F_L(q)$ and $F_H(q)$ are two independent form factors generated by
the interaction of the photon with the light quark ($d$) of charge $e_d$
and with the other heavier quark (${\bar s}$) of charge $e_{\bar s}$.
The results are then compared with predictions from vector meson dominance
(VMD) and chiral perturbation theory (ChPT). In VMD, we assume that
the photon interacts with the strange quark through the $\phi $-meson and
with the $u$-and $d$- quarks through $\rho $- and $\omega $-mesons, each
of them having the strength proportional to the quark charge. Using 
ChPT to order $p^4$, a parameter-free prediction for the $K^0$ form
factor at very low $q^2$ can be obtained which is due entirely to one-loop
diagrams without a tree-level contribution. All models are compared in
Fig.~\ref{fig:ffk0}, where the charged kaon form factor is shown for
comparison.

\begin{figure}[!ht]
\begin{minipage}[htb]{75mm}
{\psfig{figure=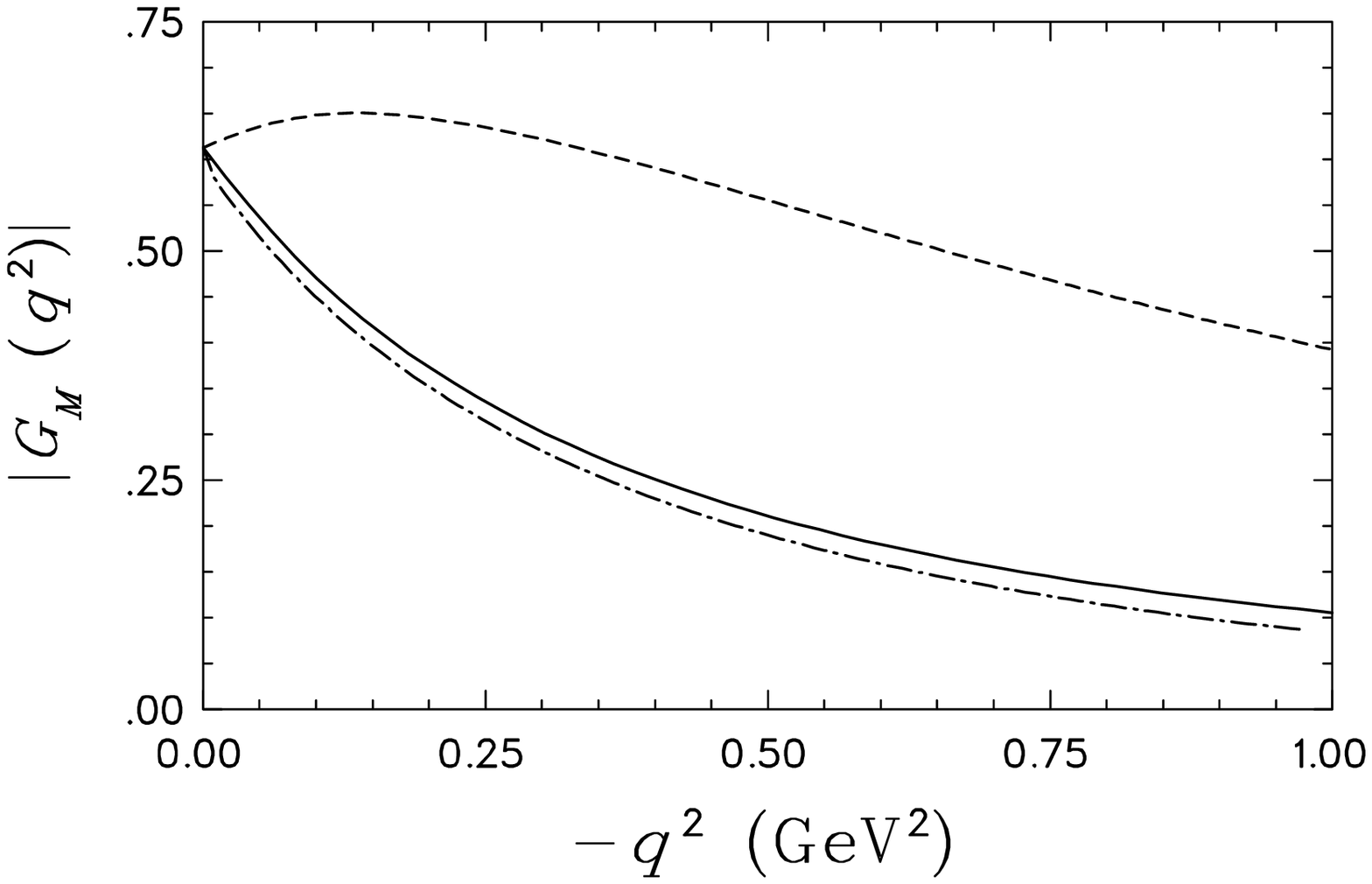,width=75mm}}
\caption{The $\Lambda$ magnetic form factors from the HVMD model (dashed line) 
 \protect\cite{williams2} and CQS model (dash-dotted line) 
\protect\cite{kim} compared with that of the neutron (solid line).
All models are scaled to the experimental $\Lambda$ magnetic moment at the
photon point.}
\label{fig:fflam}
\end{minipage}
\hspace{\fill}
\begin{minipage}[htb]{75mm}
{\psfig{figure=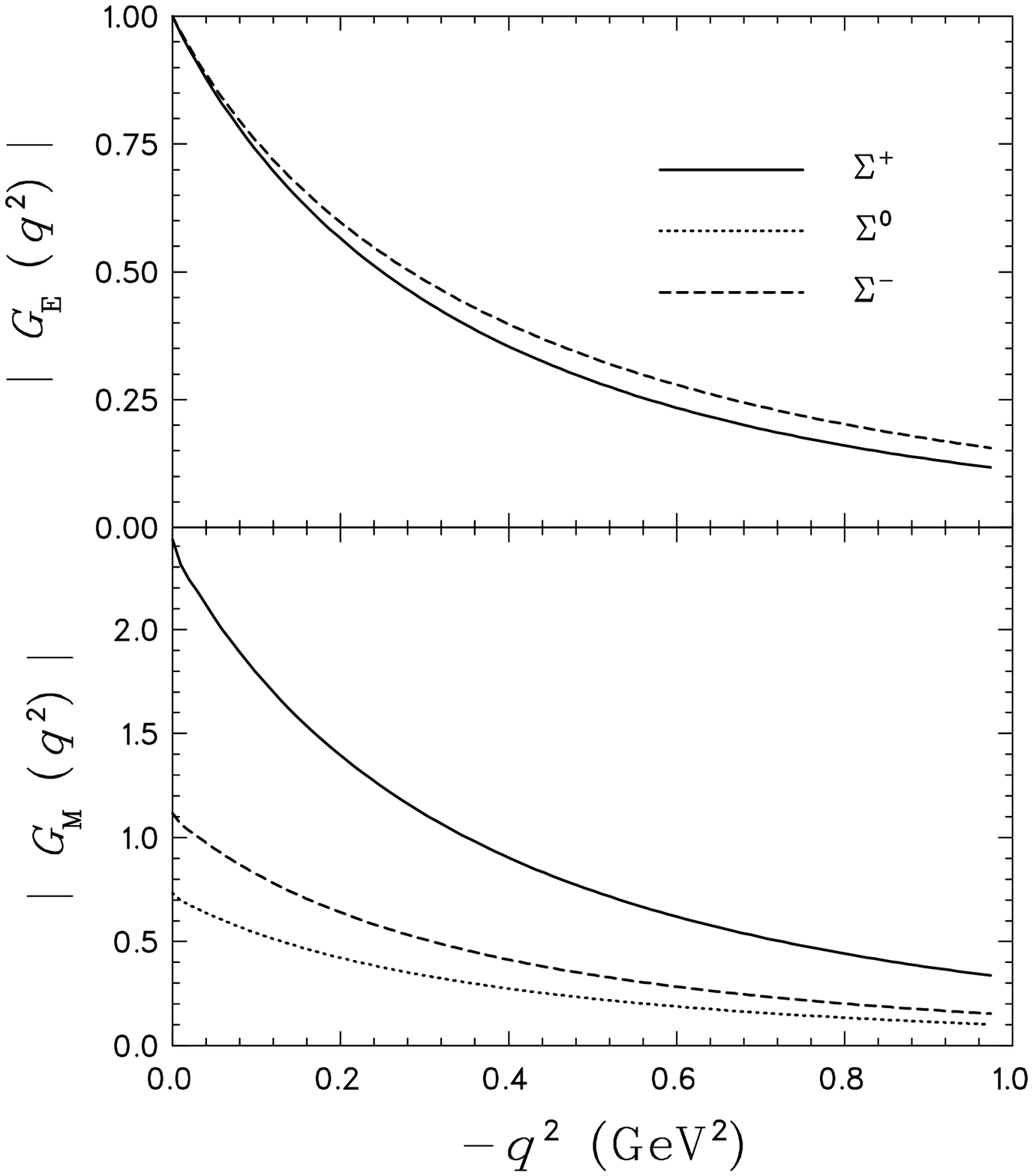,width=75mm}}
\caption{The $\Sigma$ form factors from the CQS model \protect\cite{kim}.}
\label{fig:ffsig}
\end{minipage}
\end{figure}

\subsection{The $KK^*\gamma$ Transition Form Factors}
Although several models for the $KK^*\gamma$ transition form 
factors have been introduced, no proposal for experimental verification has been suggested. 
This is in contrast to their non-strange counterparts, the $\rho \pi \gamma$ 
and $\omega\pi\gamma$ form factors, which are very important in meson exchange 
current corrections to deuteron electrodisintegration. As in the $K^0$ case, 
the transition form factors are sensitive to the mass difference between
strange and non-strange quarks. Figure \ref{fig:fftr} shows the transition 
form factors for both the neutral and the charged case, comparing the model of 
Ref.~\cite{williams} which uses vector meson dominance and the calculation of
Ref.~\cite{muenz96} which solves a covariant Salpeter equation for a confining 
plus instanton-induced interaction. Both models fall off faster than the
elastic $K^+$ form factor which is shown for comparison. The form factor
in the charged case displays a zero at $q^2=-2.7$ GeV$^2$, indicating a 
destructive interference between the light and the heavy quark contributions.

\subsection{Hyperon Form Factors}
Theoretical interest in the hyperon electromagnetic form factors is sparked 
by the question of SU(3)$_{\rm F}$ flavor symmetry breaking and the effects 
of explicit and hidden strangeness on electromagnetic observables. Applying 
SU(3)$_{\rm F}$ flavor symmetry enables predicting the hyperon form factors 
in vector meson dominance, quark and soliton models in terms of model 
parameters fixed by the nucleon data. 

In Fig. \ref{fig:fflam} we compare the $\Lambda$ form factors as predicted by a
hybrid vector meson dominance (HVMD) calculation \cite{williams2} and the
chiral quark-soliton (CQS) model \cite{kim}. The HVMD mechanism provides a 
smooth transition from the low-$q^2$ behavior predicted by vector meson 
dominance to the high-$q^2$ scaling of perturbative QCD. The key feature is 
the application of the universality limit of the vector meson hadronic 
coupling SU(3) symmetry relations. Using a direct photon coupling along with a
$\phi$ and $\omega$ pole they predict the $\Lambda$ form factor,
while the $\Lambda -\Sigma$ transition form factor can be 
obtained similarly, using a direct photon coupling and a $\rho$ pole.
Theoretically, the ratio of these form factors would be interesting since
the $\Lambda$ form factor depends only on isoscalar currents
while the $\Lambda -\Sigma$ transition depends only on isovector
contributions. Hence, as pointed out in Ref.~\cite{williams2}, this ratio
might see explicit strangeness and OZI effects such as the suppression or 
enhancement of effective $\rho$, $\omega$ and $\phi$ vector meson-hyperon 
couplings relative to the vector meson-nucleon couplings and SU(3) flavor 
symmetry predictions. 

The CQS (Nambu-Jona-Lasinio soliton) model is based on the nonlinear
effective chiral-quark model, in which the quarks interact with a
self-consistently generated background Goldstone field. The model can properly
describe the low energy pion dynamics and can be derived from QCD by means
of the instanton liquid model.

For comparison, in Fig.~\ref{fig:fflam}, we also show the neutron magnetic 
form factor with a simple, standard dipole parametrization, which were mostly
used in the previous $(e,e'K)$ studies, provided it is scaled by the $\Lambda$
magnetic moment to give the proper $\gamma\Lambda\Lambda$ vertex at the photon 
point. The
CQS model exhibits a dipole-like form factor, while the HVMD model slightly 
increases at small momentum transfers before falling off as a function of $-q^2$
for higher momentum transfers. 

Figure \ref{fig:ffsig} displays the $\Sigma$ form factors as predicted by
the CQS model \cite{kim}. We note that the charged $\Sigma$ form factors 
show a dipole behavior, as in the case of the proton. For the sake of
simplicity, we will only investigate the sensitivity of response functions
to the $\Sigma^+$ form factor.

\section{RESULTS AND DISCUSSIONS}
Table \ref{tab:result} displays the response functions that show a
sensitivity of at least 10\% to the $K^0$, $K^+K^{+*}$,
$K^0K^{0*}$, $\Lambda$, and $\Sigma^+$ form factors for certain
kinematics. No sensitivity is shown for the response functions 
involving triple polarization, since they are too 
small to be detected, about two order of magnitude smaller than the 
other response functions. Since there are many sensitive response 
functions, we will only focus on the most suitable
ones with regard to experimental aspects and proposals.

\begin{table}[!ht]
\caption{Response functions in Table 1 which are sensitive to 
         different form factors. The elementary model is due to 
         Ref.~\protect\cite{williams}. The sensitivities to the $K^0$, 
         $K^+K^{+*}$, $K^0K^{0*}$, $\Lambda$, and $\Sigma^+$ 
         form factors are denoted by $\circ$, $\bullet$, $\star$,
         $\diamond$, and $\ast$ , respectively, using the 
        $K^0 \Lambda$, $K^+ \Sigma^0$, $K^0 \Sigma^+$, $K^+ \Lambda$,
        and $K^0 \Sigma^+$ electroproduction reactions, respectively.}
\renewcommand{\arraystretch}{1.4}
\begin{center}
\begin{tabular}{|l|cc|cccccc|ccc|}
\hline\hline  
Note & $\beta$ & $\alpha$ & $T$ & $L$ & $^cTL$ & $^sTL$ & $^cTT$ & 
 $^sTT$ & $^cTL'$ & $^sTL'$ & $TT'$\\
\hline\hline  
Unpol.&0&0&$\bullet\star\diamond\ast$ & $\circ\bullet\diamond\ast$ & 
$\circ\star\diamond\ast$ &.& $\star\diamond\ast$ &. &. &. &.\\
\hline        
Pol. & 0&$x$ &.&.&.&$\ast$&.&.&$\circ\bullet\star\diamond\ast$& 
.&$\bullet\star\diamond\ast$ \\
target & 0&$y$ & $\star$ &. &$\ast$&.& . &. &.& $\circ\bullet\diamond\ast$&.\\
& 0&$z$&.&.&.&.&.& $\star$ & $\circ\bullet\star\diamond\ast$ &.& 
$\bullet\star\diamond\ast$ \\ \hline
Pol. & $x'$&0&.&.&.&.&.& $\star$&$\circ\bullet\star\diamond\ast$ 
&.&$\bullet\star\diamond\ast$ \\ 
recoil&$z'$&0&.&.&.&.&.&.&$\circ\star\diamond\ast$&.&$\bullet\star\diamond\ast$
\\ \hline
Pol. & $x'$&$x$&$\bullet\star\diamond\ast$&$\circ\bullet\diamond\ast$ & 
$\circ\diamond$ &.&.&.&.&.&.\\ 
target & $x'$&$z$& $\star\diamond$ &$\diamond\ast$&.&.&.&.&.&.&.\\
and &$z'$&$x$& $\star\diamond\ast$ &.& $\circ\bullet\diamond$ 
&.&.&.&.&.&.\\ 
recoil&$z'$&$z$&$\bullet\star\diamond\ast$&.&.&.&.&.&.&.&.\\
\hline\hline
\end{tabular}
\label{tab:result}
\end{center}
\end{table}

\begin{figure}[!ht]
\begin{minipage}[t]{75mm}
{\psfig{figure=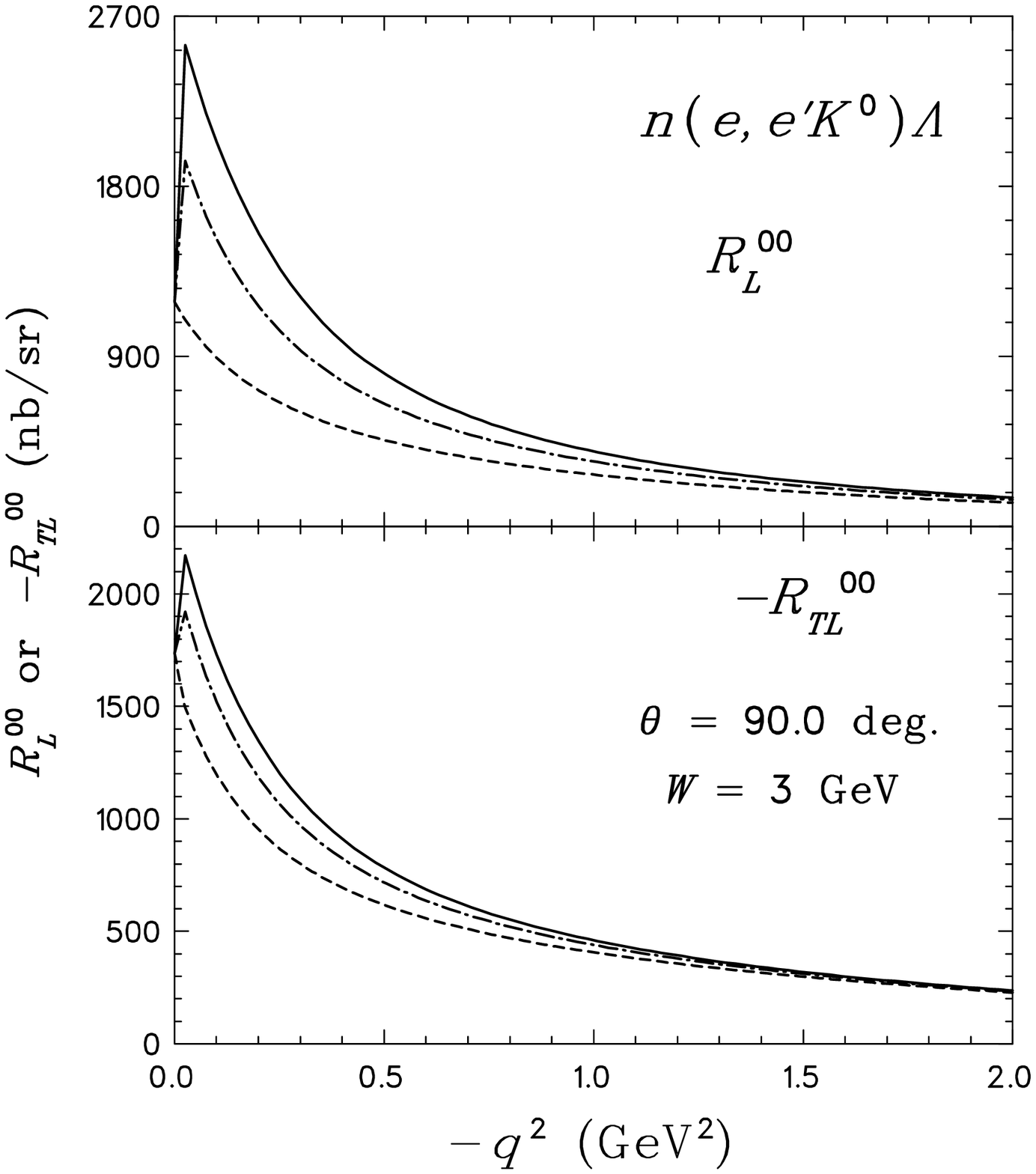,width=75mm}}
\caption{The longitudinal and longitudi- nal-transverse response functions 
 for the $n(e,e^{\prime}K^0)\Lambda$ process. The solid line
 shows the calculation with a $K^0$ form factor obtained in the LCQ model
 while the dash-dotted line was obtained using the QMV model.
 The dashed line shows a computation with the $K^0$ pole excluded.}
\label{fig:k0rfq2}
\end{minipage}
\hspace{\fill}
\begin{minipage}[t]{75mm}
{\psfig{figure=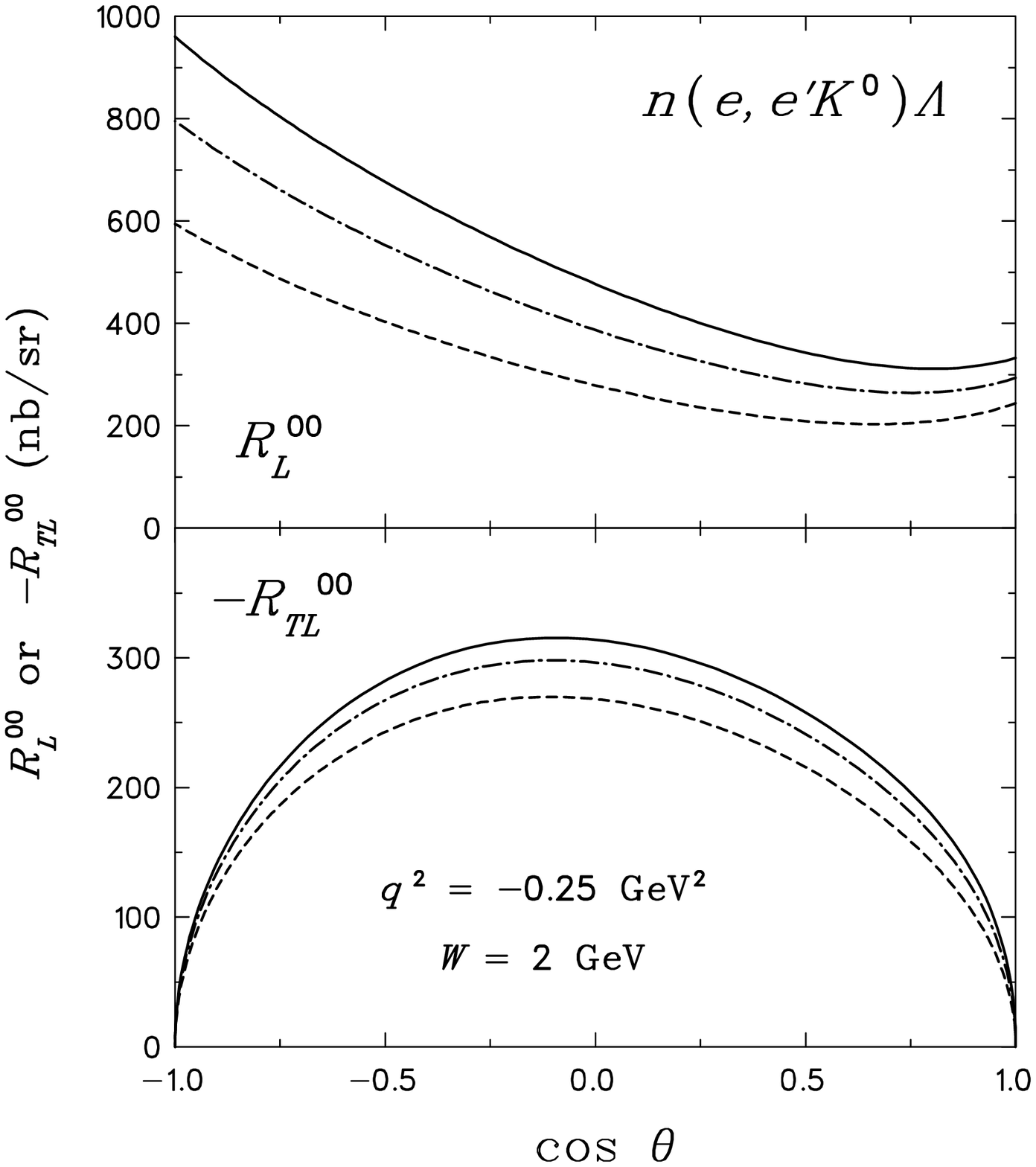,width=75mm}}
\caption{As in Fig.~\ref{fig:k0rfq2}, but for different kinematics.}
\label{fig:k0rfth}
\end{minipage}
\end{figure}

Table \ref{tab:result} can tell us, which response function is suitable
for measuring a certain form factor, while the contaminations from other
form factors can be minimized.

The longitudinal and transverse response functions of the
$n(e,e^{\prime}K^0)\Lambda$ reaction using two different quark models for
the $K^{0}$ form factor are shown in Fig.~\ref{fig:k0rfq2} and 
Fig.~\ref{fig:k0rfth}. We found that the transverse 
response functions are insensitive to the $K^0$ pole as
shown in Table \ref{tab:result}. The longitudinal response 
functions calculated with the form factor obtained from the LCQ-model is
almost 50$\%$ larger than the calculation with no $K^{0}$ pole, while the
QMV-model calculation lies between those two. Similar sensitivities of
the $L$ and $TL$ response functions indicate that a Rosenbluth separation is
not imperative in order to isolate the $K^{0}$ form factor effects.

\begin{figure}[!ht]
\begin{minipage}[t]{75mm}
{\psfig{figure=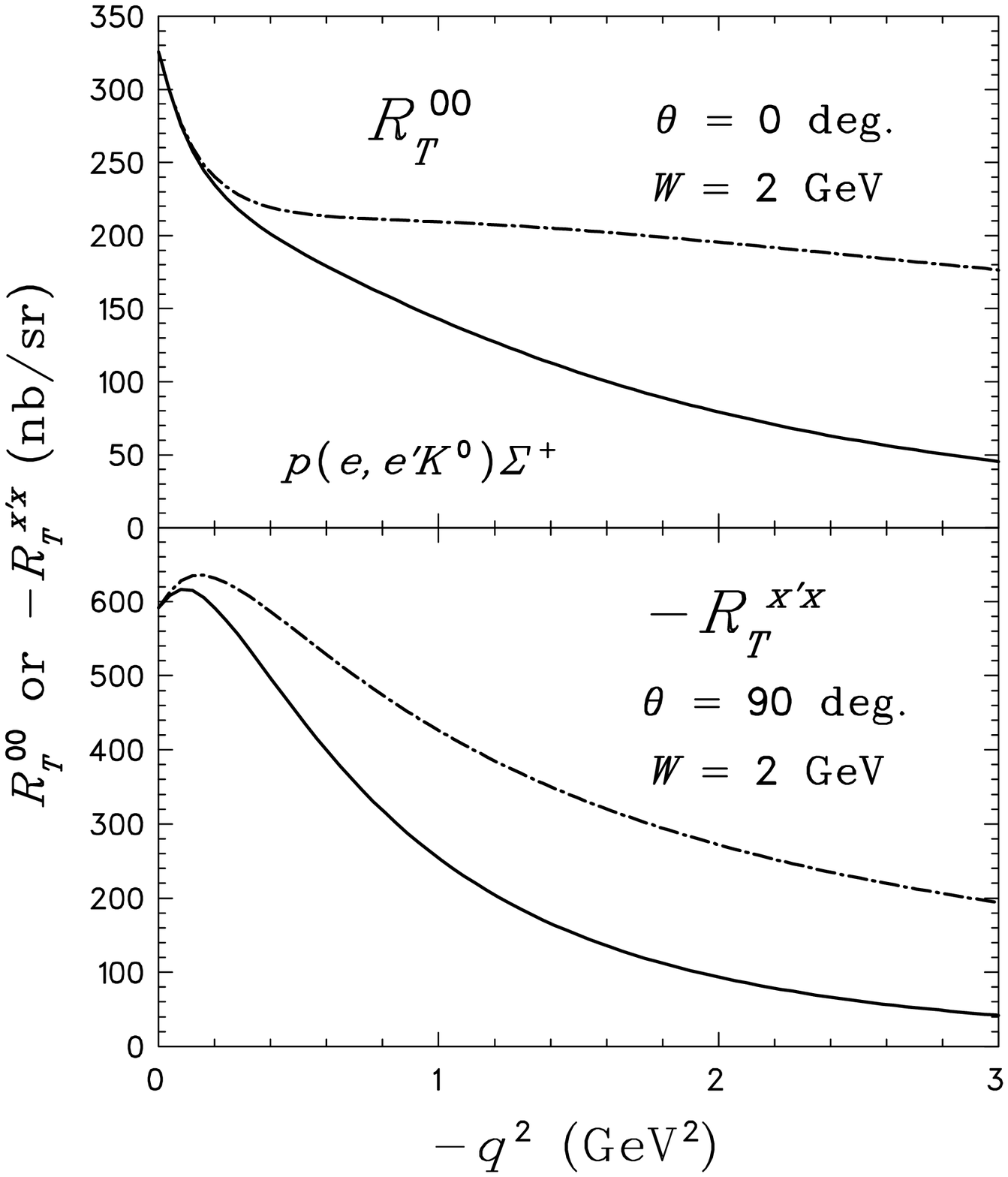,width=75mm}}
\caption{The sensitivity of response functions to different $K^0K^{*0}\gamma$
 form factors. The solid (dash-dotted) line is obtained using the model
 of Ref.~\protect\cite{muenz96} (\protect\cite{williams}).}
\label{fig:k0tr}
\end{minipage}
\hspace{\fill}
\begin{minipage}[t]{75mm}
{\psfig{figure=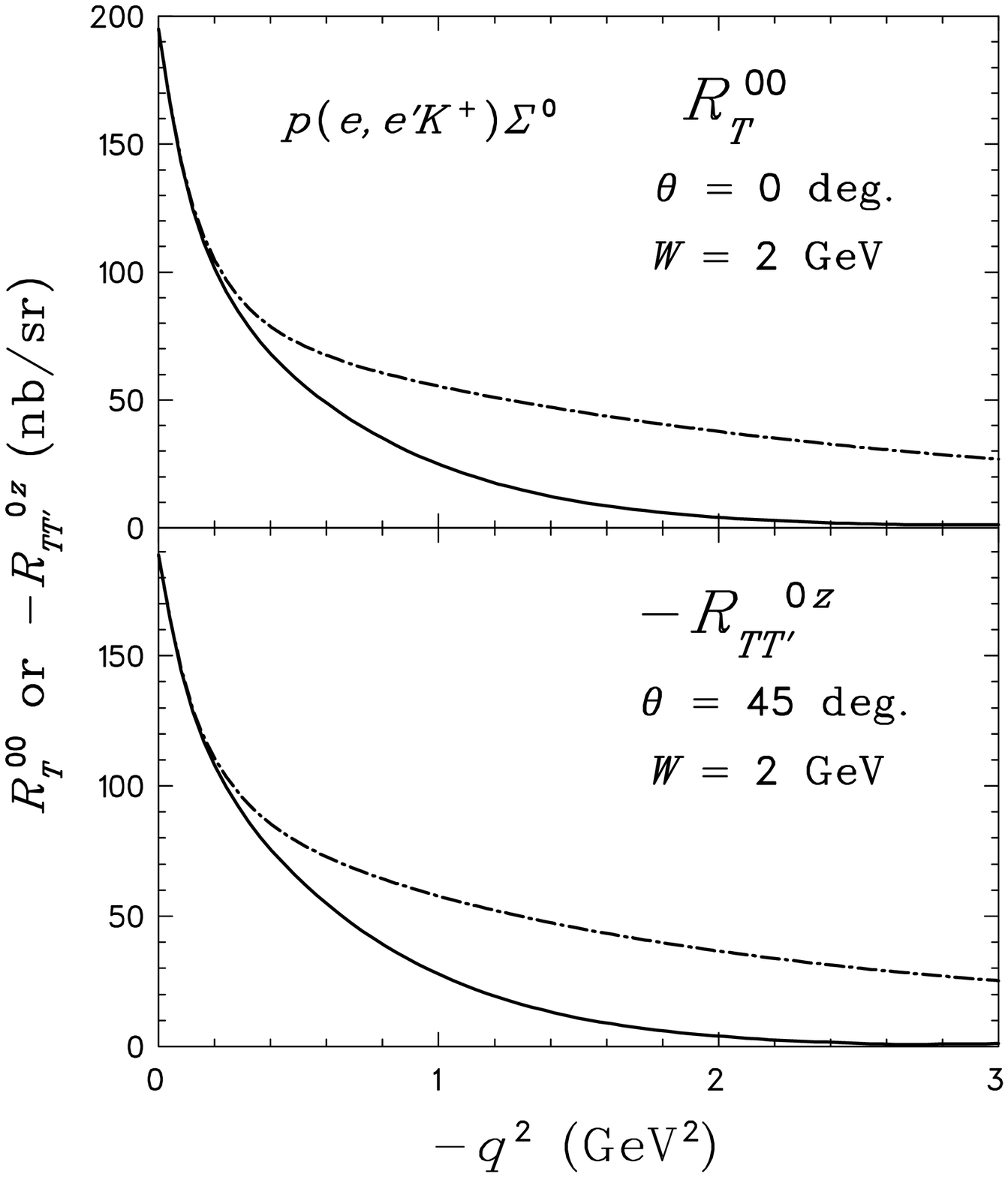,width=75mm}}
\caption{As in Fig.~\ref{fig:k0tr}, but for the $K^+K^{*+}\gamma$ form
 factors.}
\label{fig:kptr}
\end{minipage}
\end{figure}

Figures \ref{fig:k0tr} and \ref{fig:kptr} show the sensitivity of
different transverse response functions to the neutral and charged 
kaon transition form factors.
Since the small size of the $g_{K \Sigma N}$ coupling constant suppresses
the Born terms, the $p(e,e' K^+)\Sigma^0$ reaction is used to study the
$K^+ K^{*+} \gamma$ form factor while the $p(e,e' K^0)\Sigma^+$ is
sensitive to the neutral transition. Questions remain regarding additional 
$t$-channel resonance contributions from states like the $K_1$(1270) which
would have a different transition form factor. The observables displayed
can clearly distinguish between the different models, with the model of
Ref.~\cite{muenz96} leading to a much faster fall-off. Unfortunately, the
zero in the charged transition form factor around $q^2 = -2.6$ GeV$^2$ is
not visible since the response functions are already very small at this
kinematics. The large differences shown by the $T$ response function indicate
that the unpolarized experiment would be able to distinguish the models
once we had a reliable elementary production model.

\begin{figure}[!ht]
\begin{minipage}[t]{75mm}
{\psfig{figure=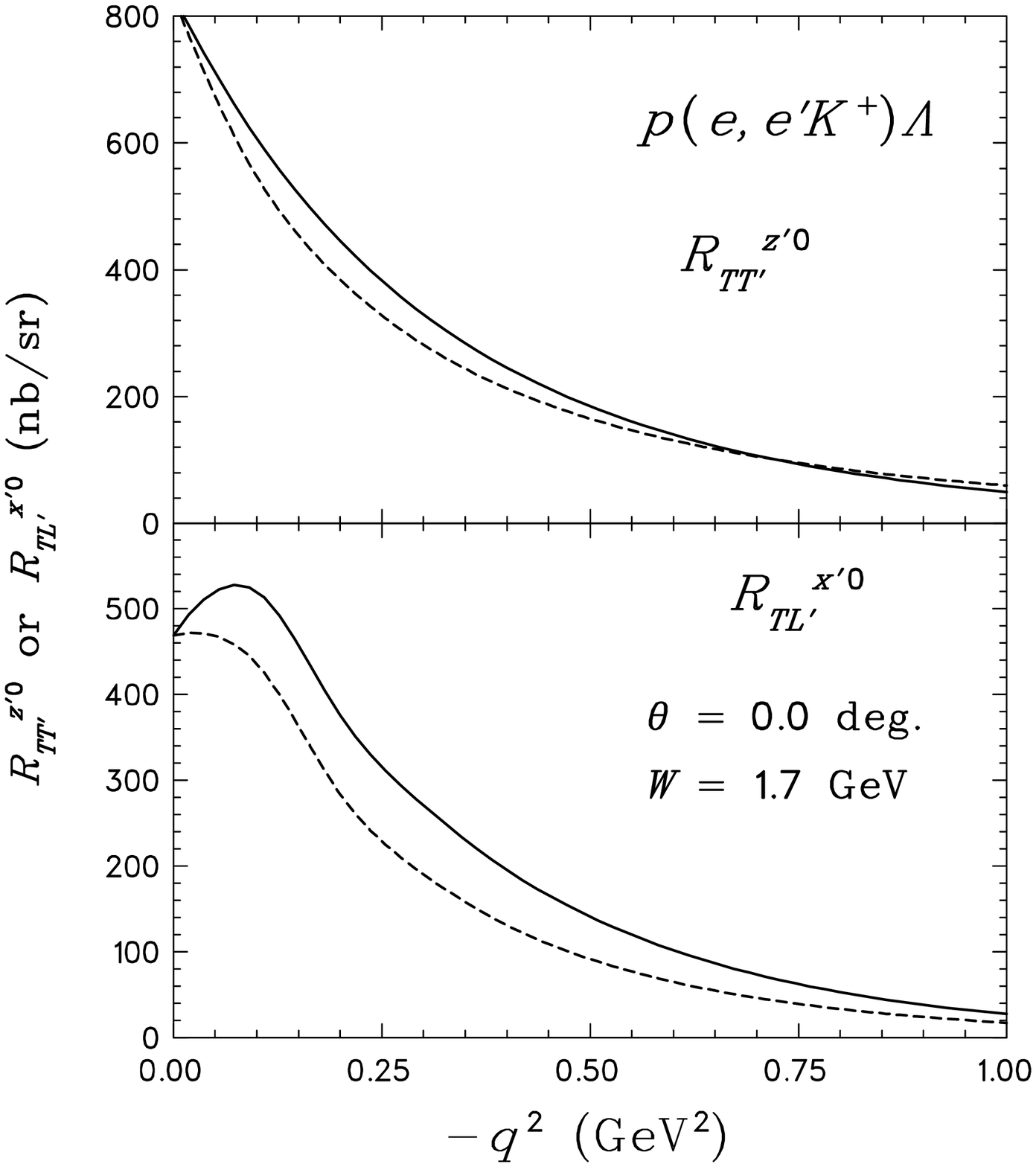,width=75mm}}
\caption{The sensitivity of response functions to different $\Lambda$
 form factors, the HVMD model (solid lines) \protect\cite{williams2} 
 and the CQS model (dashed lines) \protect\cite{kim}.}
\label{fig:lamq2}
\end{minipage}
\hspace{\fill}
\begin{minipage}[t]{75mm}
{\psfig{figure=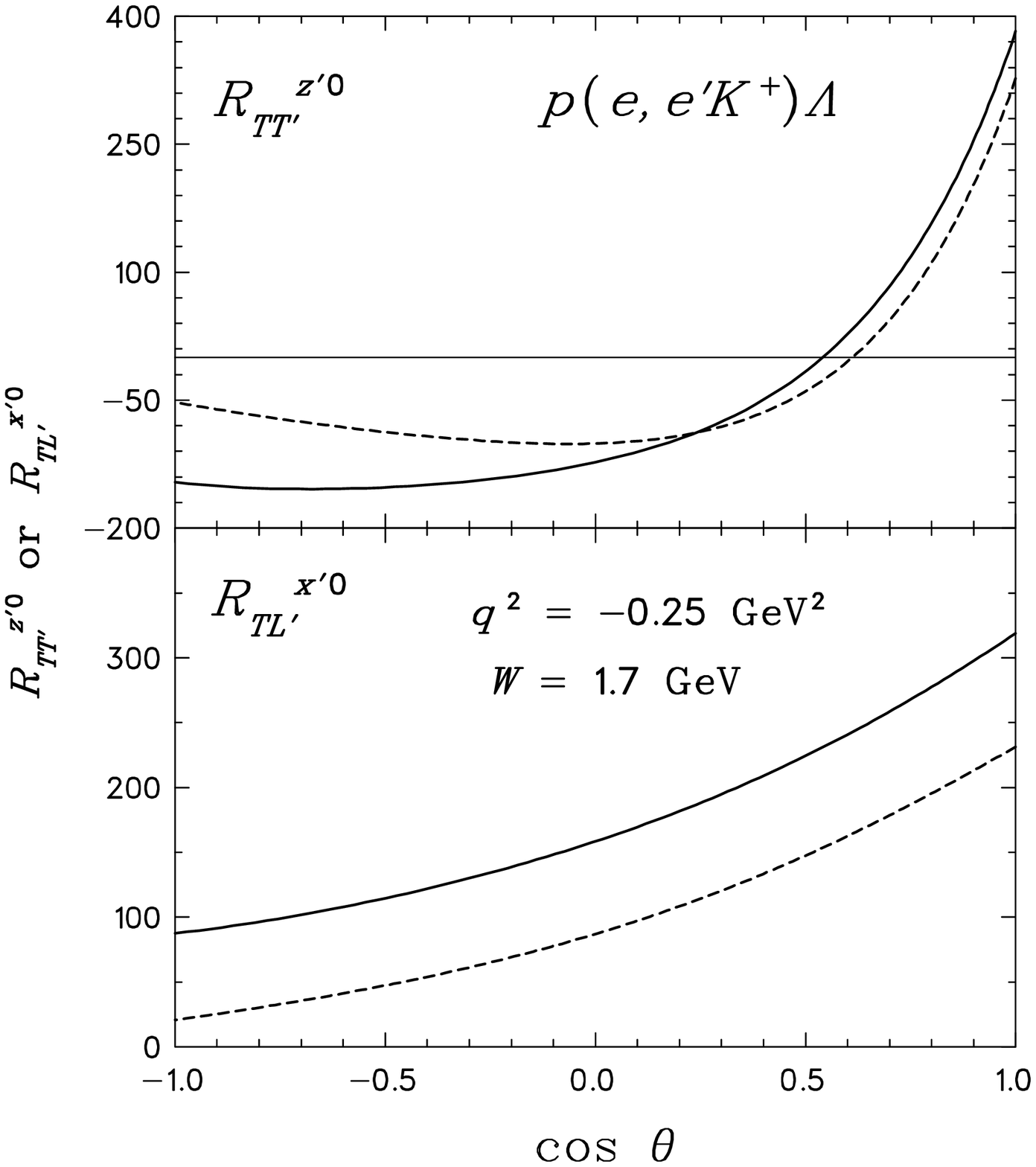,width=75mm}}
\caption{As in Fig.~\ref{fig:lamq2}, but for different kinematics.}
\label{fig:lamcth}
\end{minipage}
\end{figure}

In Fig. \ref{fig:lamq2} and \ref{fig:lamcth} we show double polarization 
response functions for the $p(e,e' K^+) \Lambda$ reaction that involve both
beam and recoil polarizations. Since the $\Lambda$ form factor 
is multiplied by the large hadronic coupling constant $g_{K\Lambda N}$, 
the $p(e,e' K^+) \Lambda$ channel is well suited to extract this form factor.
The $TT'$ and $TL'$ observables were subject of a recent TJNAF 
proposal~\cite{baker97}.  
While the $TT'$ response function shows moderate sensitivity to the different
$\Lambda$ form factors for all momentum transfers, the $TL'$
structure function displays large sensitivities for most of the momentum
transfer range shown in Fig. \ref{fig:lamq2}
and the entire angular range, shown in Fig. \ref{fig:lamcth}. The $TT'$ response shows large sensitivities to 
 the $\Lambda$ form factor for backward angles. Measuring these response 
functions could be accomplished with CLAS in TJNAF's Hall B.

\begin{figure}[!ht]
\begin{minipage}[t]{75mm}
{\psfig{figure=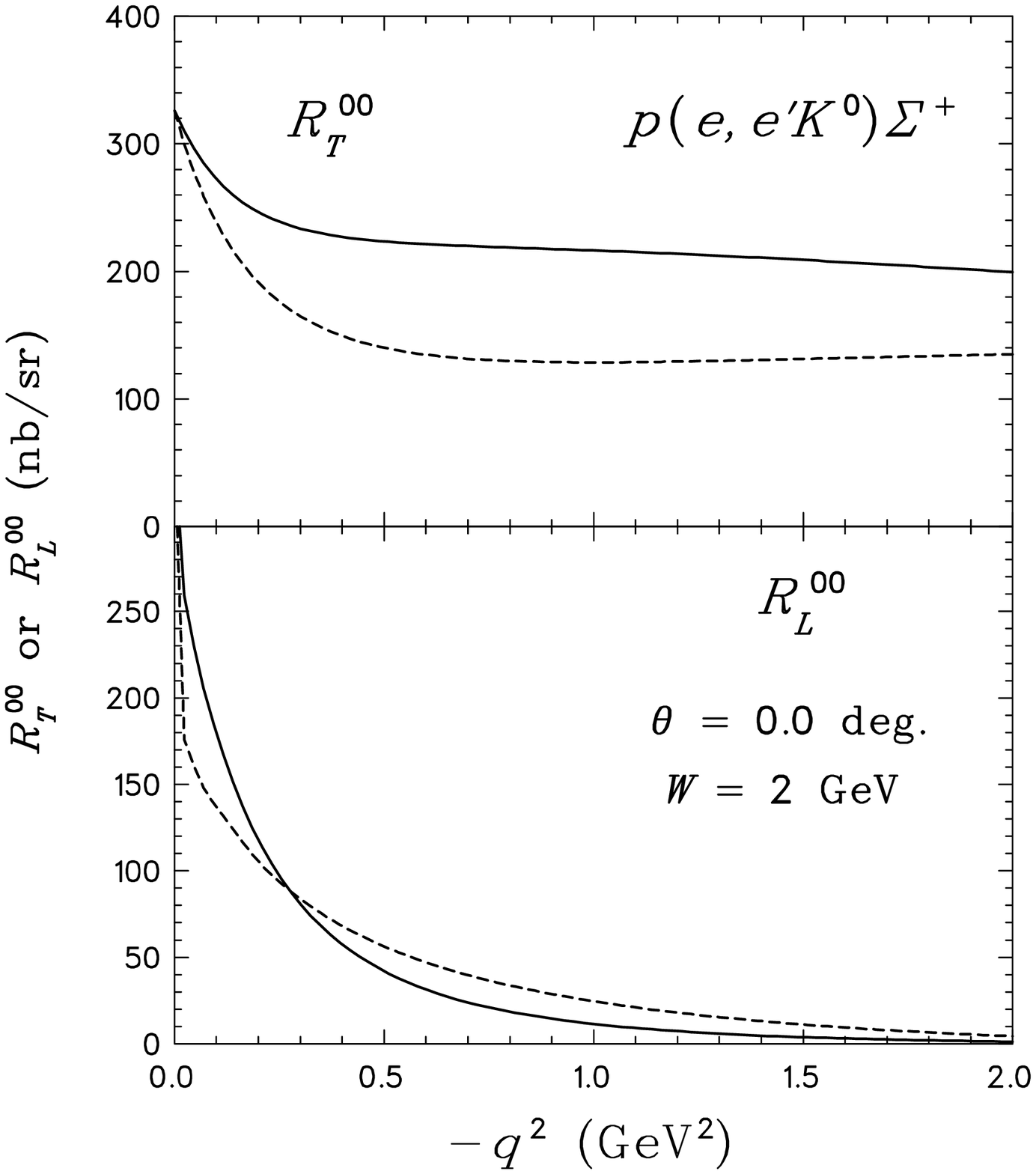,width=75mm}}
\caption{The transverse and longitudinal response functions for the 
 $K^0\Sigma^+$ channel. The dashed lines are obtained from the point $\Sigma^+$
 approximation, while the solid lines show the corresponding calculation 
 using CQS model.}
\label{fig:sigpq2}
\end{minipage}
\hspace{\fill}
\begin{minipage}[t]{75mm}
{\psfig{figure=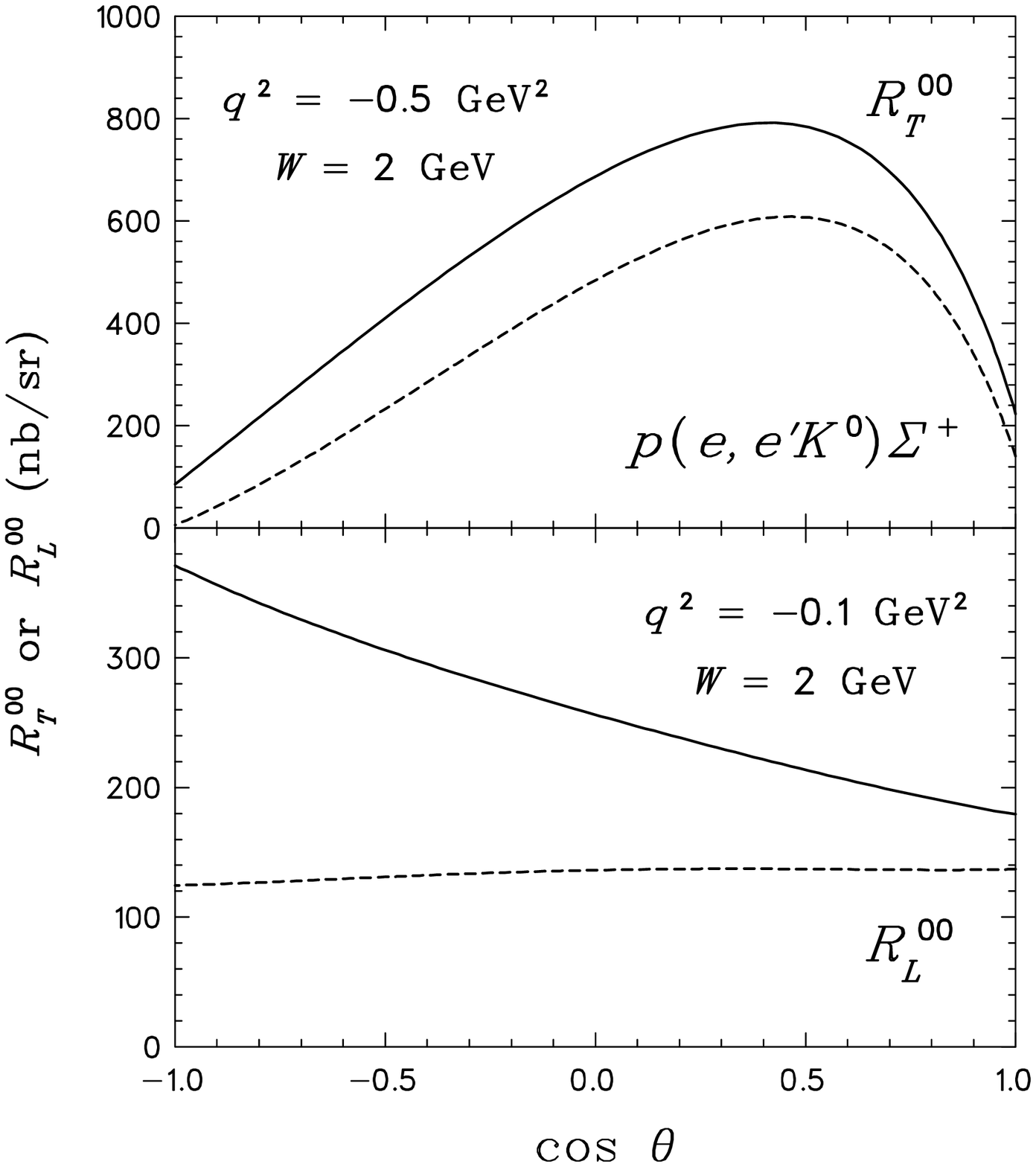,width=75mm}}
\caption{As in Fig.~\ref{fig:sigpq2}, but for different kinematics.}
\label{fig:sigpcth}
\end{minipage}
\end{figure}

Finally, in Fig.~\ref{fig:sigpq2} and \ref{fig:sigpcth} we 
show the sensitivities
of the transverse and longitudinal response functions to different 
$\Sigma^+$ form factors in the $p(e,e'K^0)\Sigma^+$ process, in which 
$\Sigma^+$ exchange is possible. Measuring this form factor has the advantage
that $\Lambda$ and $\Lambda^*$ exchanges are not possible in the $u$-channel,
therefore eliminating some uncertainties from the hyperon transition form 
factor. In previous works, we have used proton form factors, scaled
with the $\Sigma^+$ magnetic moment. We note that the CQS model predicts 
form factors which are very similar to those of the proton. Therefore, we 
compare the result only with an extreme assumption, the point $\Sigma^+$ 
approximation. Our calculations show strong sensitivities for moderate $-q^2$
and $\theta_K^{\rm cm} \geq 60^\circ$, which could easily be seen 
 experimentally. Table \ref{tab:result} indicates that
one would be able to isolate these form factors using the $R_{TL}^{0x}$ and
$R_{TL}^{0y}$ observables. However, this issue can only be addressed in
future works, since present elementary models still suffer from 
intrinsic uncertainties.

\section{CONCLUSION}
We have shown that kaon electroproduction on the nucleon is well suited to
extract information on the form factors of strange mesons and baryons, 
albeit this cannot be done model-independently, like the Chew-Low extrapolation
technique in the case of the charged pion and kaon. The main problem is to 
reduce the uncertainties in the elementary operator. This can be performed 
by precise experiments in kaon photoproduction where the relevant resonances 
and coupling constants in the process can be determined.

\section*{ACKNOWLEDGMENTS}
We thank K.S. Dhuga and O.K. Baker for useful discussions regarding the 
experimental aspects. We are grateful to H.C. Kim and K. Goeke for providing
us with their form factors. T.M. would like to thank D. Drechsel and 
L. Tiator for the hospitality during his stay in Mainz. This work is
supported by the University Research for Graduate Education (URGE) grant
and the US DOE grant no. DE-FG02-95-ER40907.


\begin{thebibliography}{99}
\bibitem{chew-low} G.F. Chew and F.E. Low, Phys. Rev. 113 (1959) 1640.
\bibitem{baker93} {\it Longitudinal/Transverse Cross Section Separation
                   for $p(e,e^{\prime }K^{+})\Lambda (\Sigma^0)$},
                   CEBAF experiment E-93-018, O.K.~Baker (spokesperson).
\bibitem{trieste97} C. Bennhold, T. Mart, and D. Kusno, {\it Proceedings of the
                    ICTP Conference on Perspectives in Hadronic Physics}, 
                    Trieste, Italy, May 1997 (in press).
\bibitem{david} J.C. David, C. Fayard, G.H. Lamot, and B. Saghai,
                Phys. Rev. C 53 (1996) 2613.
\bibitem{knoechlein} G. Kn\"ochlein, D. Drechsel, and L. Tiator, 
                     Z. Phys. A 352 (1995) 327.
\bibitem{williams} R.A. Williams, C.-R. Ji, and S.R. Cotanch, 
                   Phys. Rev. C 46 (1992) 1617.
\bibitem{terry} T. Mart, C. Bennhold, and C. E. Hyde-Wright, Phys. Rev. C
                51 (1995) R1074.
\bibitem{ito1} C. Bennhold, H. Ito, and T. Mart, {\it Proceedings of the
               7th International Conference on the Structure of Baryons},
               Santa Fe, New Mexico, 1995, p.323.
\bibitem{buck}  W.W. Buck, R. Williams, and H. Ito, Phys. Lett. B 351 (1995) 
                24; H.~Ito and F. Gross, Phys. Rev. Lett. 71 (1993) 2555.
\bibitem{williams2} R.A. Williams and T.M. Small, Phys. Rev. C 55 (1997) 882;
        R.A.~Williams and C. Puckett-Truman, Phys. Rev. C 53 (1996) 1580.
\bibitem{kim} H.C. Kim, A. Blotz, M. Polyakov, K. Goeke, Phys. Rev. D 53
              (1996) 4013.
\bibitem{muenz96} C.R. M\"unz, J. Resag, B.C. Metsch, and H.R. Petry,
                  Phys. Rev. C 52 (1995) 2110.
\bibitem{baker97} {\it Polarization Transfer in Kaon Electroproduction}, TJNAF
                Proposal, O.K.~Baker (spokesperson).
\end{thebibliography}
\end{document}